\documentclass[useAMS]{mn2e}
\usepackage{graphicx}
\usepackage[authoryear]{natbib}
\bibstyle{mnras}

\title[The UV Luminosity Function at $0.4<z<0.6$]{The ultraviolet luminosity function of star-forming galaxies between redshifts of 0.4 and 0.6}

\author[M.J. Page et al.]
{M.J. Page$^{1}$, 
T. Dwelly$^{2}$,
I. McHardy$^{3}$,
N. Seymour$^{4}$,
K.O. Mason$^{5}$,
M. Sharma$^{1}$,
\and
J.A. Kennea$^{6}$,
T.P. Sasseen$^{7}$,
A.A. Breeveld$^{1}$,
A.E. Matthews$^{1}$ 
\\ 
\\
$^{1}$Mullard Space Science Laboratory, University College London,
Holmbury St Mary, Dorking, Surrey, RH5 6NT, UK\\
$^{2}$tdastro.com, Camden Rd., Bath, BA1 5JD, UK\\
$^{3}$Department of Physics and Astronomy, University Southampton, 
Southampton SO17 1BJ, UK\\
$^{4}$International Centre for Radio Astronomy Research, Curtin University, Bentley WA 6102, Australia\\
$^{5}$Satellite Applications Catapult, Fermi Avenue, Harwell Campus, Didcot, Oxfordshire OX11 0QR, UK\\
$^{6}$Department of Astronomy and Astrophysics, The Pennsylvania State University, 525 Davey Laboratory, University Park, PA 16802, USA\\
$^{7}$Tyto Athene, 5383 Hollister Avenue, Santa Barbara, CA 93111, USA\\
}

\begin{document}

\date{Accepted ----. Received ----; in original form ----}

\pagerange{\pageref{firstpage}--\pageref{lastpage}} 
\pubyear{2023}
\maketitle

\label{firstpage}

\begin{abstract}
We combine ultraviolet imaging of the $13^{H}$ survey field, taken with the {\em XMM-Newton} Optical
Monitor telescope (XMM-OM) and the {\em Neil Gehrels Swift Observatory}
Ultraviolet and Optical Telescope (UVOT) in the UVM2 band,
to measure rest-frame ultraviolet
1,500\,\AA\ luminosity functions of star-forming galaxies with redshifts $z$
between 0.4 and 0.6. In total the UVM2 imaging covers a sky area of 
641~arcmin$^{2}$, and we detect 273 galaxies in the UVM2 image with
$0.4<z<0.6$. 
The luminosity function is fit by a Schechter function with best-fit values for
the faint end slope $\alpha=-1.8^{+0.4}_{-0.3}$ and characteristic absolute
magnitude $M^{*} = -19.1^{+0.3}_{-0.4}$. In common with XMM-OM based studies at
higher redshifts, our best-fitting value for $M^{*}$ is fainter than previous
measurements. We argue that the purging of active galactic nuclei from
the sample, facilitated by the co-spatial X-ray survey carried out with 
{\em  XMM-Newton} is important for the determination of $M^{*}$.
At the brightest absolute magnitudes ($M_{1500}<-18.5$) 
the average UV colour of our
galaxies is consistent with that of minimal-extinction local analogues, but 
the average UV colour is redder for galaxies at fainter absolute magnitudes,
suggesting that higher levels of dust attenuation enter the sample at absolute
magnitudes somewhat fainter than $M^{*}$.
\end{abstract}


\section{Introduction}
\label{sec:introduction}

Luminosity functions, the space density per unit luminosity interval as a
function of luminosity, are one of the most fundamental characterisations of
any astronomical population. Luminosity functions can be defined for
luminosities measured at any rest-frame wavelength. At optical wavelengths,
light from galaxies is contributed by stars of a variety of ages, and so the
optical luminosity function of galaxies depends on the cumulative stellar
processes in galaxies over large cosmic timescales. By contrast, at ultraviolet
wavelengths the light from massive, young stars easily overwhelms the older
stellar population, and so the luminosity is largely determined by recent
($\le$100 Myr) star-formation activity \citep{kennicutt12}, hence the
ultraviolet luminosity function of galaxies reflects the distribution of
instantaneous star formation rates.

For luminosity functions 
at ultraviolet wavelengths,
luminosity is usually described by absolute magnitude, and so the
ultraviolet luminosity function is usually defined as 
\begin{equation}
\phi = \frac{d^{2}N}{dVdM}
\label{eq:LF}
\end{equation}
where $N$ is the number of galaxies, $V$ is volume of space and $M$ is absolute
magnitude. 

UV measurements have been a key means to identify and study the star-forming
galaxy population, particularly at high redshift, since the Lyman-break
technique was employed successfully in the 1990s
\citep[e.g.][]{steidel96,madau96}. Today, rest-frame UV selection remains key
to the identification and population studies of the highest-redshift
star-forming galaxies \citep[e.g. ][]{donnan23,perezgonzalez23,harikane23}, via infrared observations with the
{\em James Webb Space Telescope}.

At low redshift, the use of ultraviolet observations to construct luminosity
functions of galaxies was pioneered by \citet{treyer98} and \citet{sullivan00},
utilising balloon-borne UV observations. They showed that the luminosity 
function is well represented by a
Schechter function shape \citep{schechter76}.
Major strides were made with the
launch of {\em GALEX} \citep{martin05}, which facilitated the construction of
UV luminosity functions all the way from $z=0$ to $z=1.2$
\citep{wyder05,arnouts05}, beyond which the rest-frame 1,500~\AA\ ultraviolet
region is sufficiently redshifted to be accessible with ground based
observations. The utility of {\em GALEX} to detect faint galaxies at
intermediate redshifts ($z>0.2$) is ultimately limited by its image resolution
 and the source confusion that becomes severe at the
faintest magnitudes \citep[$m_{NUV}>23.6$;][]{xu05}.

Both ESA's {\em XMM-Newton} observatory, and NASA's 
{\em Neil Gehrels Swift Observatory} (hereafter {\em Swift}) carry imaging UV
telescopes with better image resolution
than {\em GALEX}: the Optical Monitor
\citep[XMM-OM;][]{mason01}
and Ultraviolet and Optical
Telescope
\citep[UVOT;][]{roming05}
respectively. Deep observations with the {\em Swift} UVOT have
been used to construct UV luminosity functions between $z=0.2$ and $z=1.2$
\citep{hagen15}. Their luminosity functions were not limited by confusion, but 
colour-dependent selection effects result in faint absolute magnitude limits
similar to those of \citet{arnouts05}. More recently, \citet{page21},
\citet{sharma22} and \citet{sharma24} have used XMM-OM observations taken with
the UVW1 filter ($\lambda_{eff}=2910$~\AA) to construct rest-frame 1,500~\AA\ luminosity functions in the
redshift range $0.6<z<1.2$, with \citet{sharma22} reaching significantly
fainter absolute magnitudes than \citet{arnouts05} in this redshift
range. These studies have highlighted the importance of excluding UV-bright
active galactic nuclei (AGN) from the galaxy luminosity function to avoid
contaminating the bright end.

Very recently, \citet{bhattacharya23} have used high-resolution images from
the {\em AstroSat}
Ultraviolet Imaging Telescope \citep[UVIT; ][]{tandon17} and the {\em Hubble
  Space Telescope (HST)} to construct rest-frame 1,500~\AA\ luminosity functions in
the redshift range $0.4<z<0.8$, while \citet{sun23} used {\em HST} imaging to
construct rest-frame 1,500~\AA\ luminosity functions in
the redshift range $0.6<z<1.0$.
An earlier study by \citet{oesch10} used {\em HST}
data to construct a rest-frame 1,500~\AA\ luminosity functions in the redshift
range $0.5<z<1.0$. These {\em HST} and {\em AstroSat} studies reach fainter
absolute magnitudes than the {\em GALEX}, XMM-OM or {\em Swift} UVOT studies.

This paper is based on a UV survey of galaxies in the 13$^{H}$ Deep
Field. The 13$^{H}$ Deep Field is a patch of sky centred at
13$^{h}$~34$^{m}$~30$^{s}$~$+37^{\circ}$~53\arcmin\ (J2000) with
exceptionally low Galactic hydrogen column density,
\citep[${\rm N_{H}} \sim 7 \times 10^{19}$~cm$^{-2}$;][]{branduardi94}
and correspondingly low extinction
\citep[E(B-V)=0.005 mag;][]{schlafly11}.
It is therefore very well suited for extragalactic surveys in the UV
and soft X-ray. It was the location of the UK {\em Rosat} Deep Survey
\citep{mchardy98}
which was followed up with a long {\em XMM-Newton} exposure
\citep{loaring05} which
includes UV observations with the XMM-OM \citep{page21}.

In this paper we build on the work of \citet{page21} by again constructing 
the UV
galaxy luminosity function and examining the UV colours of galaxies, but this time in the more
recent cosmic epoch corresponding to the redshift range $0.4<z<0.6$.
UV luminosity functions have only been measured directly in this redshift
range to date in three studies \citep{arnouts05, hagen15, bhattacharya23},
of which two \citep{arnouts05, hagen15} are likely affected by AGN
contamination.
Important questions include whether the UV luminosity function changes shape
between cosmic noon and the present day, in the faint-end slope, or by
diverging from the Schechter function shape at the luminous end, and the manner
in which the UV luminosity function is shaped by extinction. At $z>2$ the
colours of UV-selected galaxies become redder with
luminosity \citep{bouwens09}, suggesting that
extinction increases with UV luminosity, but this trend may not hold towards
lower redshifts \citep{heinis13} and there is some evidence for the opposite
trend by $z=1$ \citep{sharma24}. Furthermore at $z>2$ the UV
luminosity function exhibits a steepening of the faint-end slope with redshift
\citep{reddy09,bouwens21}, but at lower redshifts, after the peak in cosmic star
formation, it is unclear if the shape is evolving.

For the redshift range $0.4<z<0.6$ we use images taken with the
UVM2 filter of XMM-OM ($\lambda_{eff}=2310$\,\AA), which is better suited
for the 
measurement of rest-frame
1,500~\AA\ luminosity than the UVW1 filter. The 13$^{H}$ Field has also been
observed through the UVM2 filter of the {\em Swift} UVOT.
The XMM-OM and UVOT have similar spatial resolution, employ similar photon-counting,
  microchannel-plate-intensified CCD detectors, and their UVM2 passband shapes
  are also similar (Fig.~\ref{fig:UVM2_passband}). Therefore, 
 to expand the sky coverage and
depth of the UVM2 imaging, we have combined the imaging from UVOT with that from
XMM-OM. 




Throughout this paper magnitudes are given in the AB system 
\citep{oke83}, and we adopt Equation~\ref{eq:LF} as our definition for the
luminosity function $\phi$. 
We have assumed cosmological parameters
$H_{0}=70$~km~s$^{-1}$~Mpc$^{-1}$, $\Omega_{\Lambda}=0.7$ and
$\Omega_{\rm m}=0.3$. Unless stated otherwise, uncertainties are
given at 1\,$\sigma$.

\section{Observations and data reduction}
\label{sec:observations}

\subsection{XMM-OM imaging}
\label{sec:xmmom}

XMM-OM observed the 13$^{H}$ field over three {\em XMM-Newton} orbits during
June 2001. During these observations the XMM-OM took seven exposures through the
UVM2 filter in Full-Frame Low-Resolution mode, each of 5000s duration. 
The observations are
listed in Table~\ref{tab:observations}.  
The XMM-OM images were initially processed with the {\em XMM-Newton} Science
Analysis System (SAS)\footnote{https://www.cosmos.esa.int/web/xmm-newton/sas}
{\sc omichain} to the stage of modulo-8 pattern noise correction.
The images were then
processed to remove the read-out streaks \citep{page17} and an
image of the UVM2 scattered-light
background structure was subtracted from each exposure to flatten the
background. The images were then distortion-corrected and re-projected in
equatorial coordinates, and astrometrically matched to objects in the Sloan
Digital Sky Survey Data Release 6 \citep{adelmanmccarthy08} using the SAS task
{\sc omatt}. The images were then summed using the SAS task {\sc ommosaic}. 

\begin{table}
\caption{Observation log for UVM2 imaging. OBSID is the observation
  identification number. Exposure time gives the total UVM2 exposure for 
each OBSID, not corrected for dead-time.}
\label{tab:observations}
\begin{tabular}{l@{\hspace{2mm}}c@{\hspace{3mm}}c@{\hspace{2mm}}c@{\hspace{2mm}}c}
\ \ OBSID & Date &\multicolumn{2}{c}{Pointing Centre} & Exposure\\
&&RA (deg)&dec (deg)&(ks)\\
\hline
&&&&\\
XMM-OM&&&&\\
0109660801&2001 Jun 12&203.665&37.913&20.00\\
0109660901&2001 Jun 22&203.665&37.913&5.00\\
0109661001&2001 Jun 24&203.665&37.913&10.00\\
&&&&\\
{\em Swift} UVOT&&&&\\
00037657002& 2008-08-12 & 203.631 & 37.787 &10.17\\
00037657003& 2008-08-13 & 203.668 & 37.773 &8.32\\
00037658001& 2011-02-17 & 203.690 & 37.750 &1.74\\
00037658002& 2011-11-04 & 203.672 & 37.782 &0.49\\
00037658003& 2011-11-09 & 203.640 & 37.796 &0.19\\
00037658004& 2011-11-13 & 203.628 & 37.795 &0.85\\
00037658005& 2011-12-07 & 203.656 & 37.773 &3.57\\
00037658006& 2012-09-03 & 203.664 & 37.809 &1.69\\
00037658007& 2012-10-17 & 203.625 & 37.804 &0.66\\
00037658009& 2013-05-18 & 203.624 & 37.713 &0.22\\
00037658010& 2013-09-03 & 203.657 & 37.780 &0.21\\
00037658011& 2013-10-17 & 203.619 & 37.796 &2.89\\
00037658012& 2013-10-19 & 203.648 & 37.793 &0.38\\
00037658013& 2013-11-01 & 203.644 & 37.795 &0.81\\
00037658014& 2013-12-07 & 203.643 & 37.782 &1.51\\
00037658015& 2013-12-12 & 203.654 & 37.776 &0.49\\
00037658016& 2013-12-20 & 203.666 & 37.773 &0.87\\
00046361001& 2014-10-18 & 203.673 & 38.043 &0.71\\
00046361002& 2019-09-06 & 203.615 & 38.035 &0.11\\
00046361003& 2019-10-25 & 203.667 & 38.050 &0.51\\
00046361004& 2019-11-02 & 203.691 & 38.056 &0.42\\
00046361005& 2019-11-03 & 203.657 & 38.035 &0.72\\
00046361006& 2021-10-17 & 203.658 & 38.068 &0.07\\
\end{tabular}
\end{table}

\subsection{{\em Swift} UVOT imaging}
\label{sec:uvot}

The 13$^{H}$ field was observed with the {\em Swift} UVOT through the UVM2
filter over 23 observations between 2008 and 2021, for a total exposure 
of 37~ks. The pointing centres were spaced such that the UVOT observations
cover a somewhat larger area than the XMM-OM observations. The observations are
listed in Table~\ref{tab:observations}. The UVOT images were
processed with a combination of {\em Swift}
ftools\footnote{https://heasarc.gsfc.nasa.gov/ftools/} and bespoke tasks. 
Initially, the raw images from each observation were corrected for modulo-8
noise using the ftool {\sc uvotmodmap}. Then the bad pixels at the corners and
right-hand edge were masked, and the images were processed to remove the
read-out streaks \citep{page13}. The images were then distortion-corrected and re-projected in
equatorial coordinates using the ftool {\sc swiftxform}. Exposure and large-scale sensitivity maps were
created using the ftools {\sc uvotexpmap} and {\sc uvotskylss}. The images were
then astrometrically matched to objects in the Sloan
Digital Sky Survey Data Release 6, and the world
coordinate systems in the exposure and large-scale sensitivity maps were
updated accordingly. 
 

\subsection{Combining the XMM-OM and UVOT images}
\label{sec:combining}

To combine the XMM-OM and UVOT images, various correction
factors which would normally be accounted for during the source
detection/photometry process were applied instead at the image stage. This
methodology is incompatible with the corrections for coincidence-loss
that are incorporated in the XMM-OM and UVOT photometry
tasks.
Coincidence-loss is the non-linearity that results from multiple
incoming photons being indistinguishable from
(and hence counted as) a single photon when they arrive in close
proximity on the detector within a single image frame \citep{fordham00}.
 However, the 
galaxies that we are
interested in are faint enough that coincidence-loss can be neglected.
 
First, the UVOT large-scale sensitivity correction was incorporated into the
UVOT exposure maps by multiplying the exposure maps by the large scale
sensitivity maps. Next the exposure maps for the XMM-OM and UVOT were divided
by the appropriate time-dependent sensitivity correction factors. Next, the
XMM-OM exposure map was divided by a factor of 2.826 to account for the
difference in UVM2 zeropoints between XMM-OM and UVOT;
see Appendix A for a full description of the origin of this factor.
Then, the background
count rates were measured in the XMM-OM and UVOT images.
The background
  count rate was found to be higher in the XMM-OM image than in UVOT. UVM2
  background
in XMM-OM and UVOT is composed of dark current, zodiacal light
and scattered light \citep{breeveld10}, and the enhanced background
in XMM-OM is dominated by
dark-current \citep{rosen23}. 
To equalise the XMM-OM and UVOT background
count rates a constant
was subtracted from the XMM-OM image.
Changing the background in this way alters the noise properties of the image
that would be inferred from the background counts, so the XMM-OM data were
then down-weighted by dividing image and
exposure by a constant factor to restore the original signal to noise
properties for faint sources assuming Poisson statistics.\footnote{
Formally, photometric measurements in XMM-OM and UVOT images are governed by
binomial statistics, but in the low count-rate limit they are equivalent to
Poisson statistics; see \citet{kuin08}.
}
The UVOT and XMM-OM
images were then summed using the ftool {\sc uvotimsum}. The exposure maps were
combined in the same way. Finally, the images and exposure maps were converted
to the format of a standard XMM-OM mosaic image. The combined UVM2 image is
shown in Fig~\ref{fig:13h_image}.


\subsection{Source detection and characterisation}

The combined XMM-OM and UVOT UVM2 image was searched for sources using the
{\em XMM-Newton} SAS task {\sc omdetect}. Photometry of sources which are
consistent with the
XMM-OM point-spread function is conducted using apertures of radius 2.8 to
5.6 arcseconds depending on brightness and the proximity of other
sources. Photometry of extended sources is obtained from connected
pixels which exceed a threshold above the background. For a more complete
description of {\sc omdetect} see \citet{page12}. In practice, the
majority of galaxies with $z>0.4$ are expected to appear point-like to XMM-OM and
UVOT. A total of 1386 sources with a signal to noise ratio $\ge4$ were detected in the UVM2 image.

\begin{figure}
\begin{center}
\includegraphics[width=50mm, angle=-90]{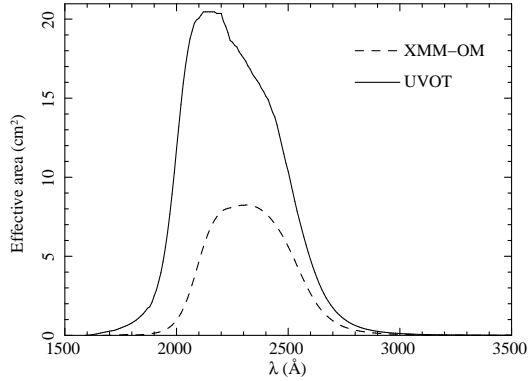}
\caption{Effective areas as a function of wavelength of the XMM-OM and {\em Swift} UVOT UVM2 passbands.}
\label{fig:UVM2_passband}
\end{center}
\end{figure}

\begin{figure*}
\begin{center}
\includegraphics[width=170mm, angle=0]{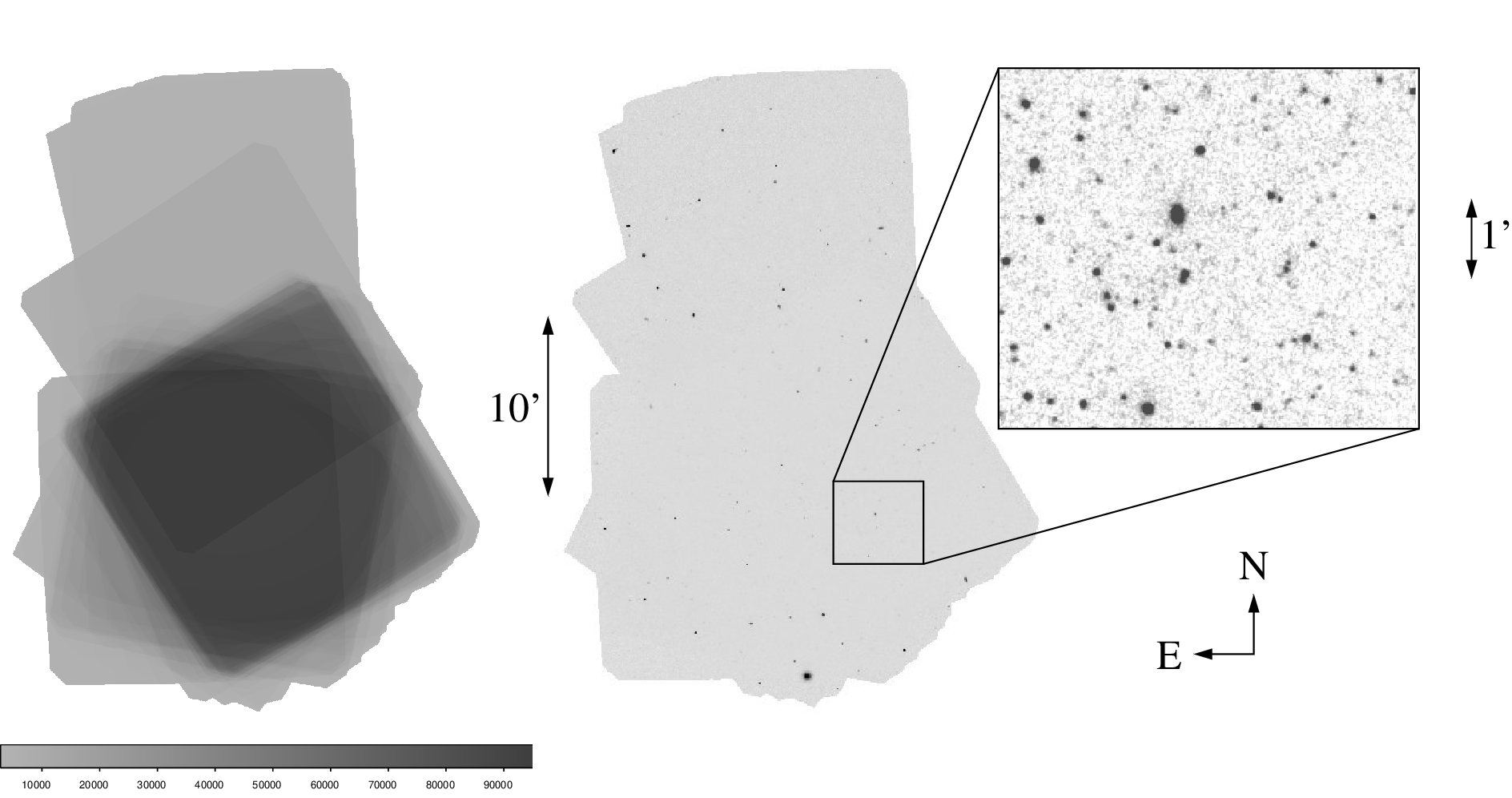}
\caption{Combined XMM-OM and UVOT UVM2 exposure map and image of the 13$^{H}$
  field. The exposure map is shown on the left; the colour bar indicates the
  exposure time in seconds. The image is shown in centre. The highlighted
  region of the image is shown at larger scale on the right to demonstrate the
  quality of the image.}
\label{fig:13h_image}
\end{center}
\end{figure*}

\subsection{Completeness}
\label{sec:completeness}

In order to construct a luminosity function, it is important to have a good
understanding of the effective sky area covered as a function of magnitude
limit, where the effective sky area is the product of the geometric sky area
and the probability of detecting a source of a given magnitude. This
probability, known as the completeness, depends on details of the source
detection method as well as the exposure time and
background levels in the image. We have calculated the completeness by
injecting fake sources at random positions in our UVM2 image and measuring the
fraction that are recovered by the source detection process. We expect the vast
majority of galaxies between redshifts of 0.4 and 0.6 to appear point-like at
the spatial resolution of XMM-OM and UVOT, hence we have injected point-like
sources. To avoid introducing artificial source confusion/crowding in the 
completeness measurements, only 20 fake sources
at a time, each with the same apparent magnitude,
are added to the UVM2
image, which is then source-searched.

A test source was considered to be recovered if {\sc omdetect} detected
a source within 2 arcsec of the input position with a signal to noise ratio 
of $\ge 4$.
The source-injection, source-search process was repeated many times
for each apparent magnitude to build up statistics on the detection
probability.
Sources were injected at UVOT magnitudes between 18 and 25 in steps
of 0.2 mag, except where the recovered fraction changes rapidly with
input magnitude ($23.4-24.0$~mag), for which the input magnitude step was
reduced to 0.1 mag. Sources were also injected with an input magnitude of 26.0.
A minimum of
1000 fake sources were injected at each input magnitude. As well as the fraction
of recovered sources, we also record the distribution of differences
between the input and recovered photometry, which represents the 
photometric error
distribution for each input magnitude. 

\begin{figure}
\begin{center}
\includegraphics[width=55mm, angle=270]{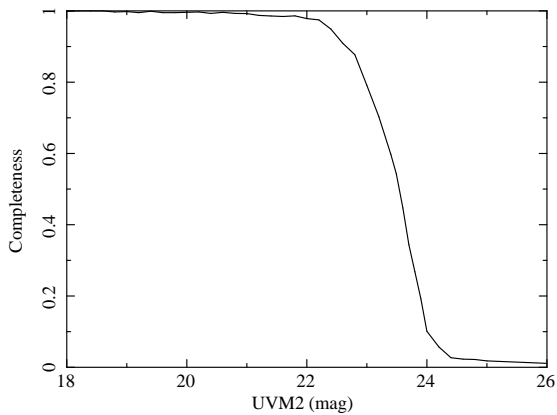}
\caption{Completeness of the source detection as a function of UVM2 magnitude,
defined as the fraction of sources, at a given input UVM2 magnitude,  which are
recovered
in the simulations described in Section~\ref{sec:completeness}.
}
\label{fig:completeness}
\end{center}
\end{figure}

Fig~\ref{fig:completeness} shows the resulting completeness, defined as the
  fraction of recovered
  sources for a specific 
input UVM2 magnitude; the curve is interpolated between the discrete input
magnitudes for which the completeness was measured.
Completeness is $> 98$ per cent for UVM2 $< 22$~mag. 
The completeness declines smoothly to 10 per cent at 24 mag. 
This rather wide
range in magnitude over which the completeness declines comes about because the
effective exposure time varies significantly across the image. At the faintest
magnitude tested with fake sources, UVM2=26~mag, one percent of injected
sources are apparently recovered. This one percent represents the combined
contributions from
otherwise-undetectable sources boosted by positive noise excursions and
source confusion in our UVM2 image.

\subsection{Redshifts}
\label{sec:redshifts}

The 13$^{H}$ Field benefits from a large number of spectroscopic redshifts and
a large body of deep optical to mid-infrared imaging from which photometric
redshifts have been derived. The spectroscopic and photometric
redshifts are described in some detail in \citet{page21}, but a brief summary
will be given here. 

Spectroscopic redshifts come primarily from multi-object spectrographs on the 
William
Herschel Telescope on La Palma, Keck and Gemini on
the Mauna Kea mountain in
Hawaii. X-ray and radio
sources were priority targets in these spectroscopic campaigns, so 
the spectroscopic campaigns have been particularly effective in identifying AGN
candidates.
Aside from X-ray and radio criteria, targetting of UV sources in spectroscopic
observations was effectively random, driven by multi-object fibre and slit
placement constraints. Of the 181 (non-X-ray, non-radio) UV sources targeted as
such in our spectroscopic observing runs, spectroscopic redshifts were obtained
for 114 (63 per cent) of the sources. Brighter sources were more likely to
yield spectroscopic redshifts: the sources for which the spectrum yielded a
redshift were on average 0.8 magnitudes brighter than those for which a
redshift was not deduced from the spectrum.
In total our campaigns have provided spectroscopic redshifts for
425 extragalactic sources in the 13$^{H}$ field.

Photometric redshifts are based on {\sc hyperz} fitting \citep{bolzonella00}, 
using up to 15
photometric bands and the spectral templates of \citet{rowan-robinson08}. 
The imaging used for photometric redshifts span the full wavelength range from the near-UV ($u^{*}$ band from
the Canada France Hawaii Telescope MegaCam) to the mid-infrared ({\em Spitzer}
$8$~$\mu$m), though the longest wavelength (5.8~$\mu$m and 8~$\mu$m) photometry
is only used in specific circumstances, as described in \citet{page21}. In
order to assign photometric redshifts, UVM2
sources were matched to the brightest $B$-band source within 2 arcsec in our
optical$-$infrared photometric catalogue. Counterparts were found for all 1,386
UVM2 sources.

Excluding broad-line AGN, there are 181 UVM2
sources with both spectroscopic and photometric redshifts. Defining the 
photometric
residuals as $\delta z = (z_{photo}-z_{spec}) / (1+z_{spec})$, where
$z_{photo}$ is the photometric redshift and $z_{spec}$ is the spectroscopic
redshift, we obtain a mean $\overline{\delta z} = -0.006\pm0.004$ and RMS
$\sigma_{\delta z} = 0.049$, for these UVM2 sources, similar to the residuals
obtained by \citet{page21} for UVW1 sources.

Spectroscopic redshifts were assigned to sources where available, with
photometric redshifts assigned to sources which did not have spectroscopic
redshifts. Of the 1,386 sources detected in the UVM2 image, 275 sources have
redshifts
between 0.4 and 0.6.

\subsection{Exclusion of AGN}
\label{sec:AGN}

It was shown in \citet{page21}, \citet{sharma22} and \citet{sharma24} that AGN
can severely distort the bright end of the UV galaxy luminosity function if
they are allowed to contaminate the galaxy sample. Fortunately, 
the 13$^{H}$ field has been
surveyed for AGN quite thoroughly, because it is also a deep X-ray and radio
survey field \citep{loaring05,seymour04}, and most of the luminous
AGN have been identified spectroscopically. 
The most serious contaminant comprises unobscured, broad-line AGN (QSOs and
type-1 Seyferts). All such objects that have been spectroscopically identified
were removed from the galaxy sample. In the redshift range of interest for this
study, this amounts to the removal of only two objects. 

As a check, we then searched for
counterparts to the remaining UVM2 sources in the X-ray catalogues of
\citet{mchardy03} and \citet{loaring05}, from {\em Chandra} and {\em
  XMM-Newton} respectively. Two sources in the $0.4<z<0.6$ redshift range are
X-ray sources, numbers 103 and 165 in the catalogue of \citet{loaring05}. They
both have X-ray luminosities greater than $10^{42}$~erg~s$^{-1}$ in the
0.5$-$7~keV band, and hence
contain AGN. However, both are spectroscopically identified through the
presence of [O\,II]~3727~\AA\ line emission and stellar features around the
Balmer break, indicative of star-forming galaxies. One (X-ray source 165) shows
significant absorption in its X-ray spectrum, but evidence for X-ray absorption
is less conclusive in X-ray source 103 \citep{page06}. As there is little
evidence for the AGN in their optical spectra, it is not possible to quantify
the extent to which their AGN may contaminate their UV emission. With absolute
magnitudes $M_{1500}$ of -19.2 and -18.5, they contribute to well-populated
bins of the UV luminosity function, and their exclusion would make little
material difference to our luminosity function measurements. We have therefore
retained these two sources in our galaxy sample.

\subsection{Source colours}
\label{sec:colours}

We derive UV colours for our sources to facilitate some investigation of their
UV continuum shapes and the degree of attenuation in their UV emission. For
this purpose we have chosen to complement our UVM2 magnitudes with u$^{*}$
magnitudes obtained from the MegaCam instrument on the Canada France Hawaii
Telescope, which are available for all but one of the UVM2-selected
sources with $0.4<z<0.6$.
Taking the effective wavelength of the u$^{*}$ passband
to be 3800~\AA, the filter is centred at a rest-frame wavelength 2533~\AA\ at
z=0.5. The u$^{*}$ images reach a 3\,$\sigma$ depth of 26.1 mag \citep{page21},
sufficient to measure the UVM2$-$u$^{*}$ colour to the faint limit of our
UVM2-selected sample.


\section{Construction of the luminosity function} 
\label{sec:construction}

\subsection{Galactic extinction}
\label{sec:extinction}

The 13$^{H}$ field is an excellent extragalactic UV survey field because it has
very small Galactic reddening and H\,I column density ($\sim 7\times
10^{19}$~cm$^{-2}$). The Galactic reddening in UVM2 was computed using the
extinction calibration from \citet{schlafly11} together with the dust map of
\citet{schlegel98}. The Galactic reddening inferred towards the 13$^{H}$ field
in the UVM2 band is 0.036 mag. The UVM2 magnitudes of our sample of galaxies,
and the magnitude limits used to compute the luminosity function, have been
corrected for this level of Galactic reddening.

\subsection{K-correction}
\label{sec:kcorrection}

\begin{figure}
\begin{center}
\includegraphics[width=55mm, angle=270]{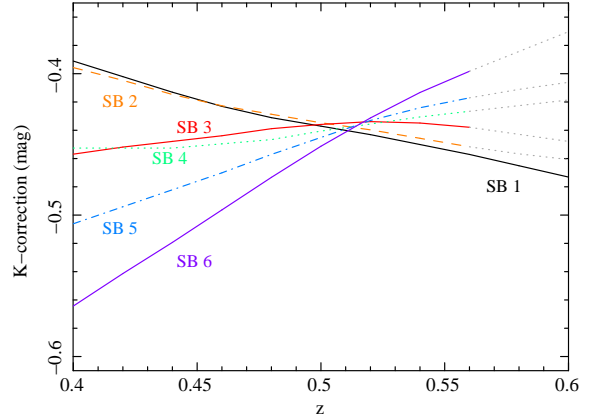}
\caption{K-corrections computed for the starburst templates of
\citet{kinney96} and \citet{calzetti94}, labelled as in
\citet{kinney96}. K-corrections for templates SB~2--6 are shown as grey dotted
lines beyond $z=0.55$
because the templates do not extend below 1250~\AA. Template
SB~1 has been extended to shorter wavelengths using the spectrum of Mrk~66 
from \citet{gonzalez98}
to permit K-corrections 
over the full redshift range.}
\label{fig:kcorr}
\end{center}
\end{figure}

K correction is the correction of photometry in the passband of observation
to a fixed rest-frame passband and is a function of redshift. For the
rest-frame passband we have opted to use the FUV passband of {\em GALEX}, which 
has a peak response close to 1,500~\AA\ and was used by \citet{wyder05} to
construct a UV galaxy luminosity function at low redshifts. The
UVM2 passband
corresponds to similar restframe wavelengths at $z=0.5$ as {\em GALEX} FUV at
$z=0$. We have calculated K-corrections for observations in the XMM-OM UVM2
passband to rest-frame {\em GALEX} FUV for the six starburst galaxy templates of
\citet{kinney96} and \citet{calzetti94}. The shortest wavelength in these
templates is 1,250~\AA, and the absence of spectrophotometry below this
wavelength begins to affect synthetic UVM2 photometry at redshifts above
0.55. Following \citet{page21}, we have extended the low-extinction SB~1
template to shorter wavelengths with the spectrum of Mrk~66 from
\citet{gonzalez98} to permit the calculation of the K-correction over the full
redshift range $0.4<z<0.6$. The resulting K corrections are shown in
Fig.~\ref{fig:kcorr}. The variation in K-corrections from the different
templates is smaller than 0.2 mag. Selection in the UV
favours low-extinction galaxies; following \citet{page21} we adopt the
K-correction curve derived from the 
low-extinction SB~1 template.

\subsection{Construction of  the binned luminosity function}
\label{sec:binnedlf}

The binned luminosity function was constructed using the method of
\citet{page00}. This involves counting the number of galaxies within an
absolute magnitude bin and dividing by the
four volume of the (volume, 
absolute-magnitude)
space
over which the galaxies were counted:
\begin{equation}
\phi_{est}(M) = \frac{N}{\int_{M_{max}}^{M_{min}}\int_{z_{min}}^{z_{max}(M)}
\frac{dV}{dz}\,dz\,dM}
\end{equation}
where $M$ is absolute magnitude, $\phi_{est}(M)$ is the binned luminosity
function, $z$ is redshift and $V$ is volume. $M_{max}$ and $M_{min}$ define the
limits of the bin in absolute magnitude, and $z_{max}(M)$ is the maximum
redshift to which an object can be detected, or the upper limit of the redshift
interval of interest (in our case $z=0.6$), whichever is the smaller.
Uncertainties on $\phi_{est}$ were calculated according to Poisson statistics
using \citet{gehrels86}.
In order
to compute the volume surveyed it is necessary to know the effective sky area
surveyed as a function of apparent UVM2 magnitude, which is the product of the
completeness and the geometric sky area. (see Section
\ref{sec:completeness}). The geometric area of our survey is 641 arcmin$^{2}$. 
For the computation of the binned luminosity function, using the completeness
vs apparent magnitude curve (Fig.~\ref{fig:completeness}), we have evaluated the
effective sky area at a set of discrete apparent magnitudes, given in
Table~\ref{tab:skyarea}. 


\begin{table}
\caption{Effective sky area as a function of magnitude, used in the construction of the binned luminosity functions.}
\label{tab:skyarea}
\begin{tabular}{cc}
UVM2 magnitude&Effective sky area \\
           (mag)&(arcmin$^{2}$)\\
\hline
&\\
20.0&638.8\\
22.2&624.7\\
22.4&608.1\\
22.6&582.4\\
22.8&561.9\\
23.0&506.8\\
23.2&450.4\\
23.4&382.5\\
23.5&347.9\\
23.6&286.4\\
23.7&219.7\\
23.8&171.7\\
\end{tabular}
\end{table}

\subsection{Measuring the Schechter function parameters}
\label{sec:fitting}

We employed a maximum-likelihood fit \citep{sandage79}
to obtain the best-fitting
Schechter-function parameters $\alpha$ and $M^{*}$. We followed the
prescription given by \citet{page21} to account for photometric uncertainty
when carrying out the fit. The approach involves minimising the following
expression:
\[
C=2N\,ln \left(\int \int \int P(M'|M,z)\phi(M)dM \frac{dV}{dz} dM' dz\right)
\]
\begin{equation}
\ \ \ \ \ \ \ \ \ \ -2\sum^{N}_{i=1} ln\, 
\int P(M_{i}'|M,z_{i})\phi(M_{i})dM
\label{eq:finalestimator}
\end{equation}
where $C$ is the quantity to be minimised, $z$ is redshift, $\phi$ is the
Schechter luminosity function, N is the number of
objects in the sample, the subscript $i$ refers to the individual sources and  
the probability of obtaining a measured 
absolute magnitude in the interval $M'$ to
$M'+dM'$ for a source of true absolute magnitude $M$ is $P(M'|M,z)dM'$.
Uncertainties on the Schechter function parameters $M^{*}$ and $\alpha$ are
obtained by varying the parameters until $\Delta C$ reaches significance 
thresholds
equivalent to those commonly applied to $\Delta \chi^{2}$ \citep{lampton76}. 
The normalisation $\phi^{*}$ is obtained by setting the model-predicted 
number of
objects in the sample equal to the observed number. 
For a full explanation/derivation of this approach see \citet{page21}.

\section{Results}
\label{sec:results}

The list of positions, redshifts, UVM2 photometry and UVM2-u$^{*}$ colours for
galaxies in the redshift range $0.4-0.6$, used to construct and fit the
luminosity function, is given in Table~\ref{tab:sourcelist}.

\begin{table*}
\caption{UVM2-selected galaxies which were used to construct the luminosity
functions. Positions are derived from the UVM2 image. UVM2 mag is the UVM2
apparent magnitude in the AB system. UVM2-u$^{*}$ is the UVM2-u$^{*}$ colour.
The column labelled $z$ gives the
redshift, and the column labelled spec/phot indicates whether
the redshift is derived from spectroscopic or photometric data. Note that
only the first five lines are included in the print version. The full table is available electronically.}
\label{tab:sourcelist}
\begin{tabular}{@{}c@{\hspace{2mm}}cccc@{\hspace{2mm}}c@{}}
\hline
RA&dec&UVM2 mag&UVM2-u$^{*}$&$z$&spec/phot\\
\multicolumn{2}{c}{(J2000)}&&&&\\
\hline
&&&&&\\
13 33 36.57 & +37 47 03.3 & $23.428\pm 0.284$ & $  0.784\pm 0.285 $ & 0.4508 &  phot \\
13 33 36.61 & +37 46 39.3 & $23.346\pm 0.220$ & $  0.506\pm 0.220 $ & 0.4948 &  phot \\
13 33 36.99 & +37 45 18.8 & $23.559\pm 0.154$ & $  0.525\pm 0.156 $ & 0.5188 &  phot \\
13 33 37.52 & +37 44 00.8 & $23.557\pm 0.180$ & $  0.353\pm 0.182 $ & 0.5995 &  phot \\
13 33 37.72 & +37 44 43.0 & $23.656\pm 0.168$ & $  0.607\pm 0.169 $ & 0.4223 &  phot \\
\end{tabular}
\end{table*}

\subsection{Source colours}
\label{sec:results_colours}

\begin{figure}
\begin{center}
\hspace{-0.5cm}
\includegraphics[width=55mm, angle=-90]{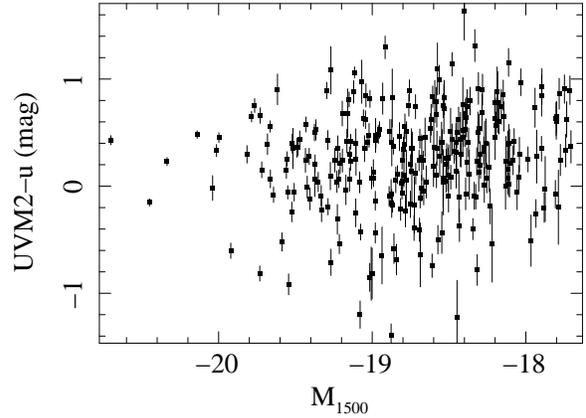}
\caption{UVM2 -- u$^{*}$ colours against absolute magnitude for the sample of
  $0.4<z<0.6$ galaxies.}
\label{fig:colours}
\end{center}
\end{figure}

\begin{figure}
\begin{center}
\hspace{-0.5cm}
\includegraphics[width=88mm, angle=0]{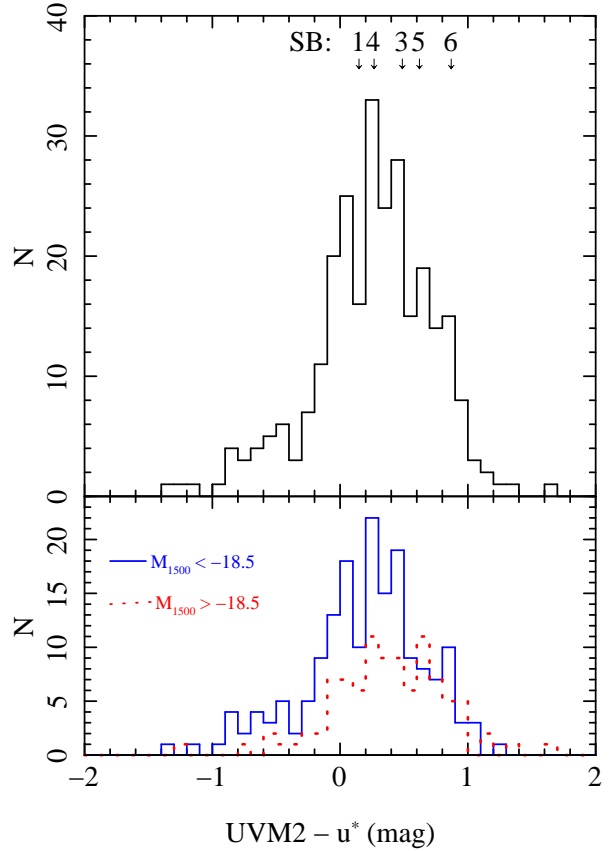}
\caption{Top panel: distribution of UVM2 -- u$^{*}$ colours for the sample of
  $0.4<z<0.6$ galaxies. At the top the colours of the starburst templates
  of \citet{kinney96} are indicated; SB~2 is not shown because it has almost
  the same colour as SB~1. Bottom panel: distribution of colours when the
  sample is split by absolute magnitude.}
\label{fig:colours_distrib}
\end{center}
\end{figure}

The UVM2$-$u$^{*}$ colours for our sources are shown against absolute magnitude in
Fig.~\ref{fig:colours}.
The colours are indicative of the UV spectral shape, which in turn is
sensitive to the level of dust attenuation \citep{meurer99}, so the
colours allow us to assess the typical levels of dust attenuation across the
UV luminosity range probed by our survey.
The distribution of colours is shown as a
histogram in Fig.~\ref{fig:colours_distrib} for the complete sample, and for
subsamples split at a threshold absolute magnitude of $M_{1500}=-18.5$. A trend
can be seen towards redder colours for the
fainter absolute magnitude subsample. The
mean colour for the full sample is $\langle$UVM2$-$u$^{*}\rangle=0.27\pm0.03$,
compared to $\langle$UVM2$-$u$^{*}\rangle=0.19\pm0.04$ for the higher luminosity
subsample and $\langle$UVM2$-$u$^{*}\rangle=0.37\pm0.05$ for the lower 
luminosity
subsample. The means of the two subsamples are different at 3$\sigma$ 
significance.


\subsection{The luminosity function}
\label{sec:results_LF}

\begin{table}
\caption{Binned luminosity function measurements. $M_{1500}$ is the centre of
  the absolute magnitude bin in the rest-frame {\em GALEX} FUV band; the
  absolute magnitude bins are 0.3 mag wide.  $\phi$ is the luminosity
  function. $N$ is the number of galaxies in each bin.}
\label{tab:magi}
\begin{tabular}{ccc}
$M_{1500}$  & Log~$\phi$ & $N$\\
   (mag)&(log [Mpc$^{-3}$~mag$^{-1}$])&\\
\hline
   $-20.72$&$-4.57^{+0.52}_{-0.76}$&   1     \\
   $-20.42$&$-4.27^{+0.37}_{-0.45}$&   2     \\
   $-20.12$&$-3.97^{+0.25}_{-0.28}$&   4     \\
   $-19.82$&$-3.66^{+0.17}_{-0.18}$&   8     \\
   $-19.52$&$-3.15^{+0.09}_{-0.10}$&  25     \\
   $-19.22$&$-2.97^{+0.08}_{-0.08}$&  35     \\
   $-18.92$&$-2.76^{+0.07}_{-0.07}$&  49     \\
   $-18.62$&$-2.61^{+0.05}_{-0.06}$&  56     \\
   $-18.32$&$-2.42^{+0.06}_{-0.06}$&  53     \\
   $-18.02$&$-2.35^{+0.09}_{-0.09}$&  28     \\
   $-17.72$&$-2.13^{+0.14}_{-0.15}$&  12     \\
\end{tabular}
\end{table}

\begin{figure}
\begin{center}
\hspace{-0.5cm}
\includegraphics[width=88mm, angle=0]{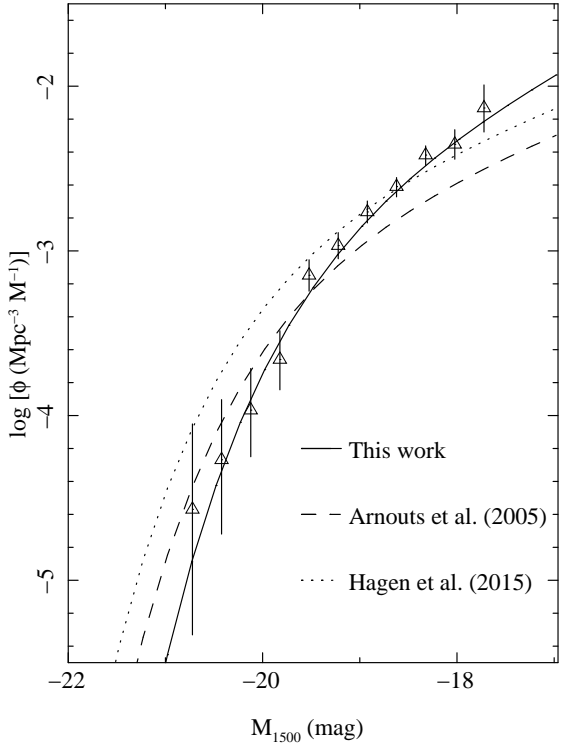}
\caption{UV luminosity function of galaxies in the redshift intervals
$0.4<z<0.6$. The datapoints show the binned luminosity function derived from
the 13$^{H}$ field as described in Section~\ref{sec:binnedlf}, and the solid
curve shows the best-fitting Schechter function derived according to the
method described in Section~\ref{sec:fitting}. For comparison, the dashed
lines show the best fitting Schechter function obtained by \citet{arnouts05},
and the dotted line shows the best-fitting maximum-likelihood Schechter
function obtained by \citet{hagen15}.}
\label{fig:magi}
\end{center}
\end{figure}

\begin{figure}
\begin{center}
\hspace{-0.5cm}
\includegraphics[width=88mm, angle=0]{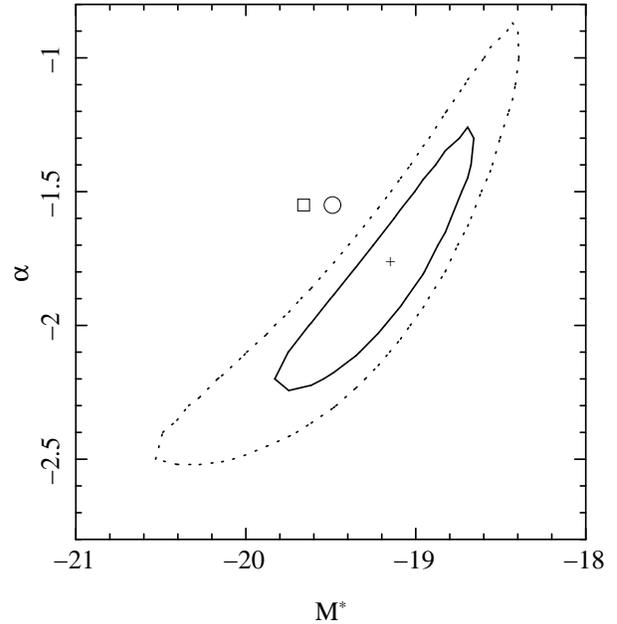}
\caption{Confidence contours for the Schechter function parameters
  $M^{*}$ and $\alpha$ from the maximum-likelihood fitting. The solid contour
  corresponds to the 68~per cent confidence region and the dotted contour
  corresponds to the 95~per cent confidence region. The cross marks the best
  fit parameter values. The circle and square mark the best fit parameter
  values found by \citet{arnouts05} and \citet{hagen15} respectively.}
\label{fig:contour}
\end{center}
\end{figure}

The binned luminosity function with a bin width of 0.3 mag is shown
in Fig. \ref{fig:magi} and tabulated in Table \ref{tab:magi}, together with the
number of galaxies contributing to each bin.  
The best-fit values for the Schechter function
parameters, with uncertainties considered independently, are
$\alpha=-1.8^{+0.4}_{-0.3}$, and $M^{*} = -19.1^{+0.3}_{-0.4}$. The confidence
contours for $\alpha$ and $M^{*}$ are shown in Fig.~\ref{fig:contour}; the
confidence contours show significant covariance between $\alpha$ and
$M^{*}$. The normalisation $\phi^{*}$, obtained as described in Section
\ref{sec:fitting} is $3.2\times 10^{-3}$~Mpc$^{-3}$. The uncertainty on
$\phi^{*}$ has contributions from the Poisson error on the number of objects in
the sample, the covariance of $\phi^{*}$ with $\alpha$ and $M^{*}$, and cosmic
variance.  The Poisson uncertainty amounts to $\pm2\times
10^{-4}$~Mpc$^{-3}$. To derive the uncertainty in $\phi^{*}$ due to its
covariance with $\alpha$ and $M^{*}$, we found the upper and lower values of
$\phi^{*}$ that correspond to the $\alpha$ and $M^{*}$ values within the
68~percent contour in Fig.~\ref{fig:contour}, which translates to an
uncertainty of $+3.9\times
10^{-3}$~Mpc$^{-3}$, $-2.2\times 10^{-3}$~Mpc$^{-3}$. The contribution from
cosmic variance, was assessed to be 15 percent, or $\pm
5\times 10^{-4}$~Mpc$^{-3}$, using the online calculator described in
\citet{trenti08}, assuming completeness and halo filling factors of 0.5, and the
\citet{sheth99} bias prescription. 
We derive an overall uncertainty for $\phi^{*}$
by adding these three contributions in quadrature, to obtain
$\phi^{*}=3.2^{+3.9}_{-2.3}\times 10^{-3}$~Mpc$^{-3}$.

\section{Discussion}
\label{sec:discussion}


\subsection{The UV colours of UV-selected galaxies}

Dust extinction is strong at UV wavelengths, and the degree to
which the UV emission from a galaxy is attenuated can be estimated from
the shape of its UV continuum \citep{meurer99}, or equivalently a UV
colour \citep{seibert05}. 
In \citet{page21} we showed that the UV colours of our $0.6<z<1.2$, UV-selected galaxies
were on average consistent with the lowest-extinction starburst template
from the sample of \citet{kinney96}. The situation is somewhat different for the
$0.4<z<0.6$ sample studied here: as seen in Section~\ref{sec:results_colours},
the average UVM2$-$u$^{*}$ colours are different for the most luminous sources
($M_{1500}<-18.5$) and the less luminous sources ($M_{1500}>-18.5$) in our
sample.
For context, a UVM2$-$u$^{*}$ colour of $-0.5$ corresponds to a spectral
  slope $\beta=-2.9$, and UVM2$-$u$^{*}=1$ corresponds to $\beta=-0.1$; 90
percent of our galaxies lie within the range delimited by these two colours.
  To facilitate comparison with the \citet{kinney96} templates, we have labelled the
positions of colours derived from the templates at $z=0.5$ on the top panel of
Fig.~\ref{fig:colours_distrib}. The template colours change little over
the redshift interval $0.4<z<0.6$: up to $\pm 0.03$ for SB~1$-$SB~4 and up 
to $\pm 0.07$ for SB~5 and SB~6. The mean colour of the luminous subsample
($\langle$UVM2$-$u$^{*}\rangle=0.19\pm0.04$) is
consistent with the expected colour (UVM2$-$u$^{*}$=0.15) of the lowest-extinction template,
SB~1. The luminosity ranges probed by \citet{page21} at $0.6<z<1.2$ do not
reach as low as $M_{1500}=-18.5$, in absolute terms, or relative to $M^{*}$ at
the respective redshifts, so could be considered equivalent to our luminous
subsample. Hence in this regard we find a consistent picture to that found at 
higher
redshift by \citet{page21}, that the most luminous UV-selected galaxies have
UV colours consistent with very low levels of extinction. 

At the lower luminosities probed by our $0.4<z<0.6$ sample ($M_{1500}>-18.5$), 
the mean colour
($\langle$UVM2$-$u$^{*}\rangle=0.37\pm0.05$) lies between the colours expected
from templates SB~3
(UVM2$-$u$^{*}$=0.49) and SB~4 (UVM2$-$u$^{*}$=0.27) indicating an increased
level of average extinction. We note that a similar pattern is seen in the UV
spectral index
measurements of UV-selected galaxies at $z\sim 1.5$ in the study of
\citet{heinis13}; in their fig.~5 the mean spectral index $\beta$ is practically
constant for all except the lowest two luminosity bins, which correspond to
$M_{1500}>-18.4$, or $M_{1500}>M^{*}+1$, in which the slope becomes redder. 
The change in mean UVM2$-$u$^{*}$ colour
between our higher and lower luminosity subsamples corresponds to a change in
spectral index $\beta$ of 0.3, similar to that observed by \citet{heinis13}. However, in other
studies focused at higher redshifts ($1<z<8$) the opposite trend
(i.e. more luminous galaxies have redder UV slopes than less luminous galaxies)
has been observed \citep{wilkins11,bouwens14,kurczynski14}.

The appearance of sources with higher levels of
extinction at absolute magnitudes somewhat fainter than $M^{*}$ is a natural
expectation if the luminosity distribution is similar for sources
with and without extinction, given the steepness of the Schechter function
shape at the bright end. Extinction will push the contribution of sources to
fainter UV absolute magnitudes, but the luminosity function rises so quickly at
the bright end as
luminosity decreases that the contribution of luminous sources with any
significant extinction will be easily overwhelmed in numerical terms by
lower-luminosity, low-extinction galaxies. Fainter than $M^{*}$, however,
the luminosity function is less steep, and it becomes possible for sources
pushed to lower absolute magnitude by extinction to contribute.   

Note that despite the small change in average colour towards fainter absolute
magnitude we have retained the K-corrections based on template SB~1 for all
galaxies in the sample. While diagnostic of the overall spectral shape, we do
not consider UVM2$-$u$^{*}$ to be a reliable predictor of the spectral shape at
$1,500$~\AA\ to the extent that we should K-correct the luminosities on an
individual basis.  The difference in K-corrections between SB~1 and SB~3 or
SB~4 amounts to only a few hundredths of a magnitude for most of the galaxies
in our sample.

\subsection{The UV luminosity function of galaxies}

Moving now to the luminosity function, we note that our survey is based on a
slightly larger sample of objects than that of \citet{arnouts05} for the same
redshift range. Our binned luminosity function (Fig.~\ref{fig:magi}) shows less
scatter in $\phi$ at the bright end than
that of \citet[][their Fig.~2]{arnouts05} but
their survey probes absolute magnitudes about 0.5 mag fainter than
ours. Compared to the survey of \citet{hagen15}, again our survey contains
slightly more objects in the redshift range $0.4<z<0.6$; our faint absolute
magnitude limit is similar to the completeness limit adopted by \citet{hagen15}
in this redshift range. The survey of \citet{bhattacharya23}, in contrast,
probes absolute
magnitudes about two magnitudes fainter than our survey, and contains twice as
many objects in the redshift range $0.4<z<0.6$, but covers only a quarter of
the sky area.

Comparing our binned luminosity function (see Fig.~\ref{fig:magi}) with the
best-fit Schechter functions of \citet{arnouts05} and \citet{hagen15}, our
luminosity function falls below their Schechter function models at the
brightest absolute magnitudes and above their models at the faintest absolute
magnitudes. Inspection of fig.~2 of \citet{arnouts05} and table~3 of
\citet{hagen15} confirms that these differences are seen directly in the binned
functions as well as in the models. Given these differences between their binned
luminosity functions and ours, it is unsurprising that their best fit models
are not compatible with ours. Comparing their best-fit values of $\alpha$ and
$M^{*}$ with the confidence intervals we derive on these parameters
(Fig.~\ref{fig:contour}), we see that they are well outside our 95 percent
confidence interval. The $\Delta C$ corresponding to their best fits (see
Section~\ref{sec:fitting})
indicates that the best-fitting model of \citet{arnouts05} is excluded by our
data at
4$\sigma$ significance, and that of \citet{hagen15} at $>5\sigma$. For
$\alpha=-1.55$, as found by \citet{arnouts05} and assumed by \citet{hagen15} in
this redshift range, our fitting would suggest $M^{*}=-18.9\pm0.1$,
0.5--0.8 magnitudes fainter than found in those studies.

That we find a fainter value of $M^{*}$ than \citet{arnouts05} or
\citet{hagen15} in the redshift range 0.4--0.6 (see Fig.~\ref{fig:mstar_alpha})
forms a consistent pattern 
with the
XMM-OM-derived results of \citet{page21}, \citet{sharma22} and \citet{sharma24}
at redshifts of 0.6--1.2. Those papers attributed the difference to the careful
purging of AGN contamination in the studies based on XMM-OM. We can investigate
the issue in our lower redshift study by considering what would happen to our
luminosity function if we put back the two AGN that were excluded in
Section~\ref{sec:AGN}. Despite this very small number of broad-line AGN, one of
them 
has an
absolute magnitude $M_{1500}=-21.25$, and would thus be the most luminous
object in the sample. Repeating the maximum likelihood fitting, with these two
AGN included, shifts the best-fit value of $M^{*}$ to -19.7, consistent with the
$M^{*}$ values found by \citet{arnouts05} and
\citet{hagen15}. We thus consider that in this study also, the different
treatment of AGN can explain the discrepancy in $M^{*}$ with the values found
by \citet{arnouts05} and
\citet{hagen15}.

\begin{figure}
\begin{center}
\hspace{-0.5cm}
\includegraphics[width=85mm, angle=0]{Mstar_alpha_vs_z_withuvithst.ps}
\caption{Direct UV measurements of Schechter function parameters $M^{*}$ and
  $\alpha$ over the redshift interval $0<z<1.2$. The points labelled `GALEX'
  correspond to the measurements of \citet{wyder05} for $z<0.1$,
  \citet{treyer05} for $0.1<z<0.2$ and \citet{arnouts05} for $z>0.2$. The
  points labelled `UVIT' correspond to the measurements of
  \citet{bhattacharya23}. The
  dashed line in the lower panel shows the result of fitting a single value to
  all measurements of $\alpha$ out to $z=1.2$.}
\label{fig:mstar_alpha}
\end{center}
\end{figure}

Fig.~\ref{fig:mstar_alpha} shows the direct measurements of the Schechter
function parameters $M^{*}$ and $\alpha$ out to $z=1$ from space-based UV data
in the literature as well as from this study. A robust trend of $M^{*}$
brightening with increasing redshift can be seen in the top panel of
Fig.~\ref{fig:mstar_alpha}, despite the scatter in $M^{*}$
values between different studies and the large variation of $M^{*}$ within the
study of \citet{bhattacharya23}.

For the faint-end slope $\alpha$, if we set aside \citet{bhattacharya23},
our measurement is one of only two covering the redshift range
$0.4<z<0.6$; our measurement is fully consistent with the other measurement
covering this redshift range, which comes from 
\citet{arnouts05}. \citet{bhattacharya23} utilised narrow redshift shells to
construct their luminosity functions, and have four separate measurements of
$\alpha$ which overlap the $0.4<z<0.6$ range. Taking the weighted average of
those four measurements gives $\alpha=-1.10\pm0.07$, flatter than our
measurement or that of \citet{arnouts05}, but consistent with both at 2$\sigma$.
Overall, there is little sign of any trend
between $\alpha$ and $z$ in Fig.~\ref{fig:mstar_alpha}. Fitting a single 
constant value to the measurements of $\alpha$ out to $z=1.2$, shown in
Fig.~\ref{fig:mstar_alpha} yields a best fit of $\alpha=-$1.29$\pm0.03$
and a
$\chi^{2}$ value of
44.3
for 24 degrees of freedom, an acceptable fit with a
null-hypothesis probability of
0.01.
On this basis we
therefore consider that the present measurements are consistent with an unchanging
$\alpha=-$1.29$\pm0.03$ over the whole redshift range $z=0$ to $z=1.2$. 
For a fixed $\alpha=-1.29$, our maximum-likelihood fitting to the 13$^{H}$
field data for $0.4<z<0.6$ would yield
$M^{*}=-18.7\pm0.1$.

In contemporary models for the galaxy population, the faint end slope $\alpha$ of the galaxy
luminosity function is primarily determined by the physics of feedback from
star-formation, particularly in driving gaseous outflows \citep{benson03,
  bower12, somerville15}. An unchanging (or little changing) faint-end slope
since $z=1.2$ would imply that the processes that regulate star formation in
low-mass galaxies have produced the same power-law distribution of star
formation rates for more than half of the Universe's history despite the
large changes in overall star-formation rate, metallicity, and
large-scale baryon disposition that have occurred during that time. Such a
robust faint-end slope would be all the more intriguing given the measurements
of a much steeper faint-end slope at earlier cosmic epochs
\citep[e.g. ][]{parsa16, mcleod24}.  

It is interesting to consider our findings on $\alpha$ in the context of two
works which examine the UV luminosity function in this redshift range, but 
are not shown in Fig.~\ref{fig:mstar_alpha} because
the UV measurements are extrapolated from longer wavelengths.
\citet{cucciati12} estimated rest-frame UV magnitudes through spectral energy
distribution model fits to deep multi-band optical and near-IR photometry for a
sample of more than 7,000 galaxies with spectroscopic redshifts. Taking the
weighted average of their estimates for $\alpha$ over the redshift interval
$0<z<1.2$ yields $\alpha=-1.05\pm0.03$, flatter than the faint-end slope 
$\alpha=-1.29\pm0.03$ derived directly from UV surveys, and inconsistent at 
6$\sigma$ significance. 

\citet{moutard20} also estimated rest-frame UV magnitudes through spectral
energy distribution fits to multi-band photometry, but using more than a
million galaxies, and predominantly with photometric redshifts. 
They included GALEX photometry in their multi-band fitting for some of
their sources, so their luminosity functions are based on a mixture of direct UV
measurements and extrapolations. \citet{moutard20} find that their fitted
values of $\alpha$ vary little with redshift. In the redshift range $0<z<1.3$
the weighted mean of their fitted values is $\alpha=-1.39\pm0.01$. This value,
in contrast to that of \citet{cucciati12}, is steeper than $\alpha=-1.29\pm0.03$ derived directly from UV surveys, and inconsistent at 
3$\sigma$ significance. Therefore in terms of the faint-end slope $\alpha$, neither
\citet{cucciati12} nor \citet{moutard20} can be reconciled with the present
ensemble of direct measurements from UV surveys,
though the discrepancy is smaller for the measurements of \citet{moutard20}.

\section{Conclusions}
\label{sec:conclusions}

We have used UVM2 imaging of the 13$^{H}$ extragalactic survey field, obtained
with XMM-OM and {\em Swift} UVOT to study the UV colours and rest-frame
1,500~\AA\ luminosity function of galaxies in the redshift range $0.4<z<0.6$.

Our luminosity function is constructed from a slightly larger sample of
galaxies than either of the comparable preceding {\em GALEX} and {\em Swift}
studies
\citep{arnouts05,hagen15} in this redshift range, and covers a larger sky area
than the {\em Astrosat} UVIT study of \citet{bhattacharya23}. We obtain a
best-fitting
Schechter function faint-end slope $\alpha=-1.8^{+0.4}_{-0.3}$, steeper but
consistent with the two previous direct determinations in this redshift range
\citep{arnouts05, bhattacharya23}.
Combining our measurement of $\alpha$ with previous
$z<1.2$ measurements from space-borne UV data, we find little evidence for any
trend with redshift, with the ensemble of measurements showing consistency with
$\alpha=-1.27\pm0.03$ at all redshifts to $z=1.2$.  Our best-fitting
characteristic magnitude $M^{*}$ is $-19.1^{+0.3}_{-0.4}$, fainter than that
found in the previous studies of \citet{arnouts05} and \citet{hagen15}. We find
that contamination of our UV-selected sample by AGN, while small in number,
would have
led to more luminous $M^{*}$ if the AGN had not been removed from the
sample. In this regard our results are in keeping with our XMM-OM based studies
at higher redshift, which also found fainter values of $M^{*}$ than previous
studies, and in which we have shown that careful purging of AGN contamination
is essential for the determination of $M^{*}$.
 
We
find that the average UV colour of the most luminous UV galaxies
($M_{1500}<-18.5$) is consistent with the lowest-extinction ($E_{B-V}<0.1$)
starburst template from the ensemble of \citet{kinney96}, implying that the
bright end of the UV luminosity function is dominated by galaxies with low
levels of dust attenuation. For absolute magnitudes fainter than
$M_{1500}=-18.5$ the average UV colour is redder, characteristic of starburst
templates with higher extinction ($E_{B-V}$ between 0.25 and 0.50),
suggesting that more dust-attenuated galaxies only start to contribute
significantly to the UV luminosity function at absolute magnitudes fainter than
$M^{*}$.

\section*{Acknowledgments}
\label{sec:acknowledgments}

Based on observations obtained with {\em XMM-Newton}, an ESA science mission
with instruments and contributions directly funded by ESA Member States and
NASA. We acknowledge the use of public data from the Swift data archive. 
This work has made use of the UK {\em Swift} Science Data Centre, hosted
at the University of Leicester, UK. MJP and AAB acknowledge support from the UK
Space Agency, grant number ST/X002055/1.

\section{Data Availability}

The primary data underlying this article are available from the 
{\em XMM-Newton} 
Science archive at https://www.cosmos.esa.int/web/xmm-newton, and the
{\em Swift}
Science data archives at https://www.swift.ac.uk and
https://swift.gsfc.nasa.gov.
Supplementary 
data underlying this article will be shared on reasonable request 
to the corresponding author.

\bibliographystyle{mn2e}

\begin{thebibliography}{}

\bibitem[Adelman-McCarthy et~al.(2008)]{adelmanmccarthy08}
Adelman-McCarthy J.K., et~al.,
2008, ApJS, 175, 297

\bibitem[Arnouts et~al.(2005)]{arnouts05}
Arnouts S., et~al., 
2005, ApJ, 619, L43

\bibitem[Benson et~al.(2003)]{benson03}
Benson A.J., Bower R.G., Frenk C.S., Lacey C.G., Baugh C.M., Cole S.,
2003, ApJ. 599, 38
  
\bibitem[Bhattacharya, Saha \& Mondal(2023)]{bhattacharya23}
Bhattacharya S., Saha K., \& Mondal C.,
2023, arXiv:2310.01903v1

%
%

\bibitem[{{Bolzonella} {et~al.}(2000){Bolzonella}, {Miralles} \&
  {Pell{\'o}}}]{bolzonella00}
{Bolzonella} M., {Miralles} J.-M., {Pell{\'o}} R., 2000, A\&A, 363, 476

\bibitem[Bower, Benson \& Crain(2012)]{bower12}
Bower R.G., Benson A.J., Crain R.A.,
2012, MNRAS, 422, 2816

\bibitem[Bouwens et~al.(2009)]{bouwens09}
Bouwens R.J., et~al.,
2009, ApJ, 705, 936  

\bibitem[Bouwens et~al.(2014)]{bouwens14}
Bouwens R.J., et~al., 
2014, ApJ, 793, 115

\bibitem[Bouwens et~al.(2021)]{bouwens21}
Bouwens R.J., et~al., 
2021, AJ, 162, 47

%
%
%

\bibitem[Breeveld et~al.(2010)]{breeveld10}
Breeveld A.A., et~al.,
2010, MNRAS, 406, 1687

\bibitem[Breeveld et~al.(2011)]{breeveld11}
Breeveld A.A., Landsman W., Holland S.T., Roming P., Kuin N.P.M., Page M.J.,
2011, AIP Conference Proceedings 1358, 373

\bibitem[Branduardi-Raymont et~al.(1994)]{branduardi94}
Branduardi-Raymont G., et~al.,
1994, MNRAS, 270, 947  

\bibitem[Calzetti, Kinney \& Storchi-Bergmann(1994)]{calzetti94}
Calzitti D., Kinney A.L., Storchi-Bergmann T., 
1994, ApJ, 429, 582

%

\bibitem[Cucciati et~al.(2012)]{cucciati12}
Cucciati O., et~al., 
2012, A\&A, 539, A31

\bibitem[Donnan et~al.(2023)]{donnan23}
Donnan C.T., et~al.,
2023, MNRAS, 518, 6011

%
%
%
%

\bibitem[Fordham, Moorhead \& Galbraith(2000)]{fordham00}
Fordham J. L. A., Moorhead C. F., Galbraith R. F.,
2000, MNRAS, 312, 83


\bibitem[Gehrels(1986)]{gehrels86}
Gehrels N., 1986, ApJ, 303, 336

\bibitem[Gonz\'alez et~al.(1998)]{gonzalez98}
Gonz\'alez R.M., Leitherer C., Heckman T., Lowenthal J.D., Ferguson H.C., 
Robert C., 
1998, ApJ, 495, 698

\bibitem[Hagen et~al.(2015)]{hagen15}
Hagen L.M.Z., Hoversten E.A., Gronwall C., Wolf C., Siegel M.H., Page M., 
Hagen A.,
2015, ApJ, 808, 178

\bibitem[Harikane et~al.(2023)]{harikane23}
Harikane Y., et~al., 
2023, ApJS, 265, 5

\bibitem[Heinis et~al.(2013)]{heinis13}
Heinis, S., et~al., 
2013, MNRAS, 429, 1113

%
%
%
%

\bibitem[Kennicutt \& Evans(2012)]{kennicutt12}
Kennicutt R.C. \& Evans N.J., 
2012, ARA\&A, 50, 531

\bibitem[Kinney et~al.(1996)]{kinney96}
Kinney A.L., Calzetti D., Bohlin R.C., McQuade K., Storchi-Bergmann T., 
Schmitt H.R., 
1996, ApJ, 467, 38

\bibitem[Kurczynski et~al.(2014)]{kurczynski14}
Kurczynski P., et~al., 
2014, ApJ, 793, L5

\bibitem[Kuin \& Rosen(2008)]{kuin08}
Kuin N.P.M. \& Rosen S.R.,
2008, MNRAS, 383, 383

\bibitem[Lampton, Margon \& Bowyer(1976)]{lampton76} 
Lampton M., Margon B., Bowyer S., 
1976, ApJ, 208, 177

\bibitem[Loaring et~al.(2005)]{loaring05}
Loaring N.S., et~al., 
2005, MNRAS, 362, 1371

\bibitem[Madau et~al.(1996)]{madau96}
Madau P., Ferguson H.C., Dickinson M.E., Giavalisco M., Steidel C.C., 
Fruchter A.,
1996, MNRAS, 283, 1388

%
%

\bibitem[Martin et al.(2005)]{martin05}
Martin D.C., et al., 
2005, ApJ, 619, L1

\bibitem[Mason et~al.(2001)]{mason01}
Mason K. O., et~al.,
2001, A\&A, 365, L36

\bibitem[McLeod et~al.,(2024)]{mcleod24}
McLeod D.J., et~al.,
2024, MNRAS, 527, 5004

\bibitem[McHardy et~al.(1998)]{mchardy98}
McHardy I.M., et~al., 
1998, MNRAS, 295, 641

\bibitem[McHardy et~al.(2003)]{mchardy03}
McHardy I.M., et~al., 
2003, MNRAS, 342, 802

\bibitem[Meurer, Heckman \& Calzetti(1999)]{meurer99}
Meurer G.R., Heckman T.M. \& Calzetti D.,
1999, ApJ, 521, 64

%
%
%

\bibitem[Moutard et~al.(2020)]{moutard20}
Moutard T., Sawicki M., Arnouts S., Golob A., Coupon J., Ilbert O., 
Yang X., Gwyn S.,
2020, MNRAS, 494, 1894


\bibitem[Oesch et.~al.(2010)]{oesch10}
Oesch P.A., et~al., 
2010, ApJ, 725, L150

\bibitem[{{Oke} \& {Gunn}(1983)}]{oke83}
{Oke} J.~B., {Gunn} J.~E., 1983, ApJ, 266, 713

\bibitem[Page \& Carrera(2000)]{page00}
Page M.J., Carrera F.J., 
2000, MNRAS, 311, 433

\bibitem[Page et~al.(2006)]{page06}
Page M.J., et~al.,
2006, MNRAS, 369, 156

\bibitem[Page et~al.(2012)]{page12}
Page M.J., et~al.,
2012, MNRAS, 426, 903

\bibitem[Page et~al.(2013)]{page13}
Page M.J., et~al., 
2013, MNRAS, 436, 1684

\bibitem[Page et~al.(2017)]{page17}
Page M.J., et~al.,
2017, MNRAS, 466, 1061

\bibitem[Page et~al.(2021)]{page21}
Page M.J., et~al., 
2021, MNRAS, 506, 473

\bibitem[Parsa et al.(2016)]{parsa16}
Parsa S., Dunlop J.S., McLure R.J., Mortlock A.,
2016, MNRAS, 456, 3194

\bibitem[P\'erez-Gonz\'alez et~al.(2023)]{perezgonzalez23}
P\'erez-Gonz\'alez P.G., et~al., 
2023, arXiv:2302.02429

\bibitem[Poole et~al.(2008)]{poole08}
Poole T.S., et~al.,
2008, MNRAS, 383, 627

\bibitem[Reddy \& Steidel(2009)]{reddy09}
Reddy N.A., Steidel C.C.,
2009, ApJ, 692, 778

\bibitem[Roming et~al.(2005)]{roming05}
Roming P.W.A., et~al.,
2005, Space Sci. Rev., 120, 95

\bibitem[{{Rowan-Robinson} {et~al.}(2008){Rowan-Robinson}, {Babbedge},
  {Oliver}, {Trichas}, {Berta}, {Lonsdale}, {Smith}, {Shupe}, {Surace},
  {Arnouts}, {Ilbert}, {Le F{\'e}vre}, {Afonso-Luis}, {Perez-Fournon},
  {Hatziminaoglou}, {Polletta}, {Farrah} \& {Vaccari}}]{rowan-robinson08}
{Rowan-Robinson} M., {et~al.}, 2008, MNRAS, 386, 697

\bibitem[Rosen et~al.(2023)]{rosen23}
Rosen S. \& OMCal Team,
2023, XMM-SOC-CAL-TN-0019 Issue 8.0, available from
https://www.cosmos.esa.int/web/xmm-newton/calibration-documentation  

\bibitem[Sandage, Tammann \& Yahil(1979)]{sandage79}
  Sandage A., Tammann G.A., Yahil A.,
  1979, ApJ, 232, 352

\bibitem[Schlafly \& Finkbeiner(2011)]{schlafly11}
Schlafly E.F., Finkbeiner D.P., 
2011, ApJ, 737, 103

\bibitem[Schechter(1976)]{schechter76}
Schechter P., 
1976, ApJ, 203, 297

\bibitem[Schlegel, Finkbeiner \& Davis(1998)]{schlegel98}
Schlegel D.J., Finkbeiner D.P., Davis M., 
1998, ApJ, 500, 525

\bibitem[Seymour, McHardy \& Gunn(2004)]{seymour04}
Seymour N., McHardy I.M. \& Gunn K.F., 
2004, MNRAS, 352, 131 

%
%
%
%

\bibitem[Sharma, Page \& Breeveld(2022)]{sharma22}
Sharma M., Page M.J. \& Breeveld A.A.,
2022, MNRAS, 511, 4882 

\bibitem[Sharma et~al.(2024)]{sharma24}
Sharma M., Page M.J., Ferreras I., Breeveld A.A.,
2024, MNRAS, 531, 2040

\bibitem[Sheth \& Tormen(1999)]{sheth99}
Sheth R.K., \& Tormen G., 1999, MNRAS, 308, 119

\bibitem[Seibert et~al.(2005)]{seibert05}
Seibert M., et~al., 2005, ApJ, 619, L55

\bibitem[Somerville \& Dav\'e(2015)]{somerville15}
Somerville R.S. \& Dav\'e R.,
2015, ARA\&A, 53, 51

\bibitem[Steidel et~al.(2006)]{steidel96}
Steidel C.C., Giavalisco M., Pettini M., Dickinson M., Adelberger K.L., 
2006, ApJ, 462, L17

\bibitem[Sullivan et~al.(2000)]{sullivan00}
Sullivan M., Treyer M.A., Ellis R.S., Bridges T.J., Milliard B., Donas J.,
2000, MNRAS, 312, 442

\bibitem[Sun et~al.(2023)]{sun23}
Sun L., et~al.,
2023, arXiv:2311.15664v1

%

\bibitem[Tandon et~al.(2017)]{tandon17}
Tandon S.N., et~al.,
2017, JApA, 38, 28  

\bibitem[Trenti \& Stiavelli(2008)]{trenti08}
Trenti M., \& Stiavelli M., 
2008, ApJ, 676, 767

\bibitem[Treyer et~al.(1998)]{treyer98}
Treyer M.A., Ellis R.S., Milliard B., Donas J., Bridges T.J.,
1998, MNRAS, 300, 303

\bibitem[Treyer et~al.(2005)]{treyer05}
Treyer M., et~al.,
2005, ApJ, 619, L19

%

\bibitem[Wilkins et~al.(2011)]{wilkins11}
Wilkins S.M., Bunker A.J., Stanway E., Lorenzoni S., Caruana J., 
2011, MNRAS, 417, 717 

\bibitem[Wyder et~al.(2005)]{wyder05}
Wyder T.K., et~al.,
2005, ApJ, 619, L15

\bibitem[Xu et~al.(2005)]{xu05}
Xu C.K., et~al.,
2005, ApJ, 619, L11

\end{thebibliography}

\newpage
\appendix
\vspace{3mm}
\noindent
{\bf APPENDIX A: Relative throughput of XMM-OM and {\em Swift} UVOT employing
  the UVM2 filter}
\vspace{1mm}

\noindent
In order to use a combined UVOT and XMM-OM image to generate our catalogue of
UVM2-selected galaxies, the count rates in the data from one of the instruments
must be scaled so that a source of a given UVM2 magnitude will correspond to the same count
rate in both instruments. We have implimented this adjustment by
scaling the XMM-OM exposure map. The scaling is based on the difference in
zeropoints for the two instruments, but also must take into account the
different aperture corrections that relate the instrumental
zeropoints to the
measurement aperture employed in the source-searching and measurement software
(in our case, {\sc omdetect}). The scaling must also account for colour terms
that arise from the small difference in the shapes of the UVM2 passband between
the two instruments. We will describe each of these components in turn.
\vspace{2mm}

\noindent
{\bf APPENDIX A.1 instrumental zeropoints}

\noindent
The AB zeropoint for XMM-OM in the UVM2 filter is 17.412,
for a 17.5 arcsec radius aperture \citep{rosen23}. The AB zeropoint for Swift UVOT in the UVM2 filter is 18.54,
for a 5 arcsec radius aperture \citep{breeveld11}. 
\vspace{2mm}

\noindent
{\bf APPENDIX A.2 aperture corrections}

\noindent
The photometric measurements in our UVM2 source catalogue are obtained from the
software {\sc omdetect}. Almost all galaxies at $0.4<z<0.6$ are
indistinguishable from point sources at the spatial resolution and depth of our UVM2
image. For faint, point-like sources, {\sc omdetect} employs a circular
aperture of radius 2.8 arcsec to measure photometry \citep{page12}, hence for
both XMM-OM and UVOT the different aperture corrections from the aperture for
which their instrumental zeropoint is defined to the 2.8 arcsec radius
measurement aperture must be
taken into account.

The XMM-OM has a stable PSF thanks to its carefully controlled thermal
environment. The aperture correction from the 17.5 arcsec aperture that
corresponds to the instrumental zeropoint to a 2.8 arcsec aperture is therefore obtained from the
{\em XMM-Newton} Current Calibration
Files\footnote{https://www.cosmos.esa.int/web/xmm-newton/current-calibration-files}. This
aperture correction is 0.278 magnitudes.

On the other hand, the PSF of the UVOT shows small variations as the thermal
environment of the telescope changes over Swift's 90 minute orbit. Hence aperture
corrections for UVOT are best measured directly from the UVOT images
on which they will be employed \citep{poole08, breeveld10}. Therefore we
measured aperture corrections from nine stars of appropriate magnitude in the
UVOT image by obtaining, for each star, photometry in 5 arcsec and 2.8
arcsec apertures using the {\em Swift} FTOOL {\sc uvotsource}. For each star the difference between the magnitudes obtained in the two apertures
represents a measurement of the aperture correction. These measurements
were then averaged to obtain a mean aperture correction of $0.249\pm0.007$
magnitudes from a 5 arcsec
radius aperture (for which the instrumental zeropoint is defined) to the 2.8
arcsec measurement aperture.  

The difference between the XMM-OM and UVOT aperture corrections is 0.029
magnitudes.
\vspace{2mm}

\noindent
{\bf APPENDIX A.3 colour correction}

\noindent
As seen in Fig.~\ref{fig:UVM2_passband}, the shapes of the UVM2 bandpass are
similar but not identical for XMM-OM and UVOT. Therefore in comparing the
photometric responses of the two instruments, there is a small colour term to
consider, which will depend on the spectrum of the source being observed.
We have synthesized these colour terms for the six template galaxy spectra
which were used to investigate the K-correction in
Section~\ref{sec:kcorrection}. The colour term can be expressed as UVM2$_{\rm XMM-OM}-$UVM2$_{\rm UVOT}$
where UVM2$_{\rm XMM-OM}$ is the magnitude of an object in the UVM2 bandpass of
XMM-OM and UVM2$_{\rm UVOT}$ is the magnitude of an object in the UVM2 bandpass of
UVOT. The colour terms range from $-0.01$, $-0.02$ and $-0.04$ magnitudes for
the lowest extinction SB~1, SB~2, and SB~3 templates respectively at $z=0.4$ to
$-0.06$, $-0.06$ and
$-0.08$ magnitudes for the highest extinction SB~4, SB~5 and SB~6 templates at
$z=0.5$, with the variation of colour with redshift contributing up to 0.02
magnitudes. Given that our sample is dominated by low extinction sources (see
Section~\ref{sec:results_colours}) we consider that a typical colour term is
UVM2$_{\rm XMM-OM}-$UVM2$_{\rm UVOT}=-0.03$.
\vspace{2mm}

\noindent
{\bf APPENDIX A.3 overall throughput adjustment}

\noindent
The difference in instrumental UVM2 zeropoints between XMM-OM and {\em Swift} UVOT is 
1.128 mag. To this we should add an aperture correction of 0.029 mag and a
colour correction of $-0.03$ magnitudes, but as the aperture and colour
corrections almost perfectly cancel, we have simply adopted the difference in the
instrumental zeropoints to represent the throughput difference.
\vspace{2mm}

\noindent
{\bf APPENDIX A.4 systematic errors}

\noindent
It is important to consider the degree of systematic error we may be
introducing by combining the XMM-OM and UVOT images.

The systematic uncertainty on the instrumental zeropoint (i.e. the uncertainty
on global photometry) for {\em Swift} UVOT is reckoned to be 0.03 mag
\citep{poole08}. There is no equivalent number in the calibration documentation
for XMM-OM, but as the two instruments are calibrated against overlapping sets
of photometric standard stars, we expect any error in the zeropoint of UVOT to
be replicated systematically in the XMM-OM zeropoint. Hence we would expect the
zeropoint uncertainty of 0.03 mag to apply similarly to the combination of
XMM-OM and UVOT data.

Uncertainties arising from the aperture corrections would be expected to be
dominated by the uncertainty on the UVOT aperture correction, i.e. 0.007
mag. Uncertainties related to colour correction, given the spread in
corrections for different galaxy templates, could amount to as much as 0.03
mag, much larger than the uncertainty due to aperture effects. The quadrature
sum of the uncertaintes from aperture and colour corrections is 0.03 mag,
similar to the level of systematic uncertainty on the zeropoints of the two
instruments.

For our study of faint galaxies, we have taken a signal to noise threshold of
4. This corresponds to a statistical photometric uncertainty of 0.3~mag.  This
statistical uncertainty is around ten times larger than the systematic
uncertainty that might be contributed by combining XMM-OM and UVOT data, and
hence our photometric uncertainty is dominated by statistical, rather than
systematic, uncertainty.
\label{lastpage}

\end{document}